\documentclass{appolb}
\usepackage{graphicx}
\usepackage{amsmath}
\begin{document}
\title{Some remarks on non-planar Feynman diagrams%
}
\author{Krzysztof Bielas\thanks{Presented by K. Bielas at the International Conference of Theoretical Physics ``Matter To The Deepest'', Ustron 2013}, Ievgen Dubovyk, Janusz Gluza
\address{Institute of Physics, University of Silesia,\\ Uniwersytecka 4, PL-40-007 Katowice, Poland}
\\ \vspace{0.5cm}
{Tord Riemann
}
\address{Deutsches Elektronen-Synchrotron, DESY,\\ Platanenallee 6, 15738 Zeuthen, Germany}
}

%\author{Krzysztof Bielas\thanks{Presented by K. Bielas at the International Conference of Theoretical Physics ``Matter To The Deepest'', Ustron 2013}, Ievgen Dubovyk, Janusz Gluza}
%\address{Institute of Physics, University of Silesia, Uniwersytecka 4, PL-40-007 Katowice, Poland}\\
%\author{Tord Riemann}
%\address{Deutsches Elektronen-Synchrotron, DESY, Platanenallee 6, 15738 Zeuthen, Germany}

\maketitle
\begin{abstract}
Two criteria for planarity of a Feynman diagram upon its propagators (momentum flows) are presented.
Instructive \texttt{Mathematica} programs that solve the problem and examples are provided. A simple geometric
argument is used to show that while one can planarize non-planar graphs by embedding them on higher-genus
surfaces (in the example it is a torus), there is still a problem with defining appropriate dual
variables since the corresponding faces of the graph are absorbed by torus generators.
\end{abstract}
  
\section{Introduction}

Non-planar Feynman diagrams arise naturally from perturbative quantum field theory. They are interesting for
many reasons. First of all, from the graph-theoretical point of view many constructions and theorems are
formulated only for planar graphs. Formally, any Feynman diagram $G$ can be considered as a graph and thus
subjected to graph-theoretical methods. Moreover, labelling all edges of $G$ by momenta makes $G$ a network
flow, and which is not necessarily unique. Consequently, it can make some redundant problems on the way to
find effective analytical or numerical solutions to a given Feynman diagram. This is especially true if we
want to make general programs which use some methods to solve Feynman integrals. We focus here on one
technical aspect. Given a Feynman diagram $G$, is it possible to decide the planarity of $G$ only upon its
propagators? The question could be rewritten in a more general way: is it possible to decide the planarity of
a network flow only upon its flows? The answer is important considering e.g. computer algebra methods in
particle physics. We 
faced the problem when working on the upgrade of \texttt{AMBRE} package \cite{ambre}, where different methods
are applied in order to construct an optimal, low-dimensional Mellin--Barnes representation of $G$
% (i.e. the smallest dimensions of Mellin--Barnes integrals) 
depending on its planarity (work in progress).
The problem of planarity identification of $G$ upon its propagators has been mentioned lately also in
\cite{lee}. In general, the available graph-theoretical methods to recognize planarity of a graph rely mainly
on its geometry, like the Kuratowski theorem, that claims that a graph $G$ is planar iff it does not contain a
subgraph that is a subdivision of $K_{3,3}$ or $K_5$ \cite{Nakanishi}. When no geometry of $G$ is given,
it is hard to decide about subgraphs of $G$. This is the case of \texttt{AMBRE}, where only propagators
are given. As we will show, the answer for the question is positive.

Another interesting property of non-planar Feynman diagrams is that they are not dealt with twistor methods,
originally applied only to planar sector of $\mathcal{N}=4$ SYM \cite{Arkani}. The question whether it is
possible to apply these methods to the non-planar sector remains open. The main obstacle is the lack of duals
for non-planar diagrams, hence lack of dual variables, on which these methods rely. On the other hand, in the
so-called 't Hooft limit of $\mathrm{SU}\left(N\right)$ with coupling $g$, where $N\rightarrow\infty$, $g^2
N=\mathrm{const}$, only planar diagrams survive. Thus one could argue that (non-)planarity is not of purely
technical, graph theoretical character, but rather it is a significant ingredient with a physical
interpretation in the above limit.

%The organization of the paper is as follows. In further part of introduction some elements of graph theory are introduced. In the section \ref{sec:lapl} the combinatorial method is presented. In the section \ref{sec:dual} the geometrical method is presented. In the section \ref{sec:twist} a simple geometrical consideration about dual variables for non-planar diagrams is presented.

Let us start with some definitions. A graph is \textit{planar} if it can be drawn on a surface (sphere) without intersections. A \textit{non-planar} graph is a graph that is not planar.

%The simplest non-planar graphs are Kuratowski graphs:
%\begin{figure}[htb]
%\centerline{%
%\includegraphics[scale=0.4]{nonpl}}
%\caption{$K_5$ (left) and $K_{3,3}$ (right)}
%\label{kur}
%\end{figure}

\begin{figure}[bth]
\centerline{%
\includegraphics[scale=0.55]{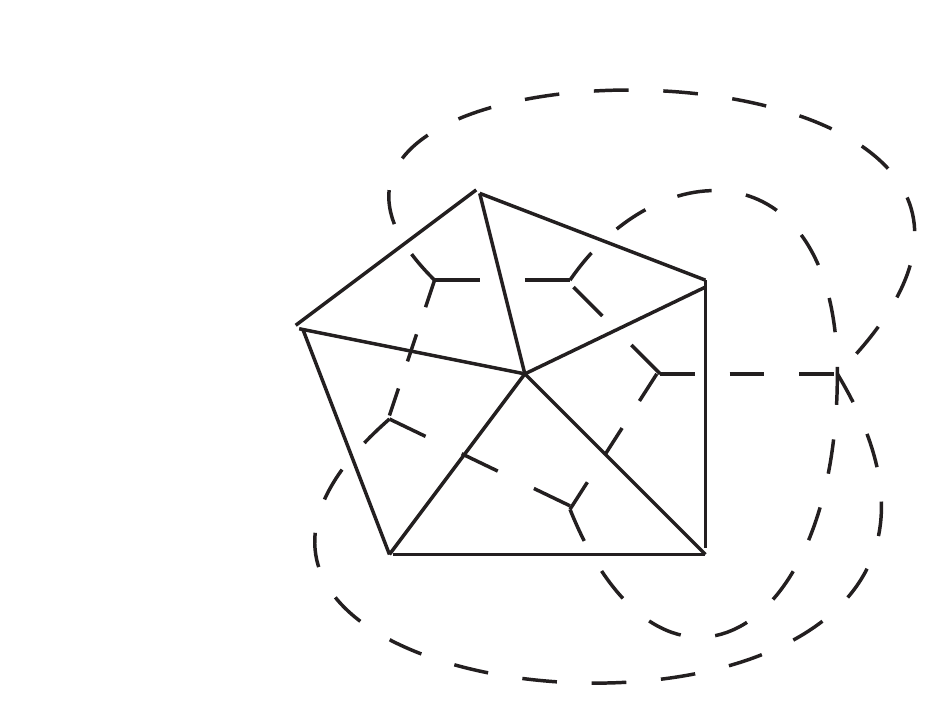}}
\caption{Given graph (solid line) and its dual (dashed line)}
\label{du1}
\end{figure}

A \textit{dual} to a graph is constructed by drawing vertices inside the faces (including the external face) and connecting vertices that correspond to adjacent faces (Fig. \ref{du1}). 

Such duals can be defined only for planar graphs.

\begin{figure}[t]
\centerline{%
\includegraphics[scale=0.7]{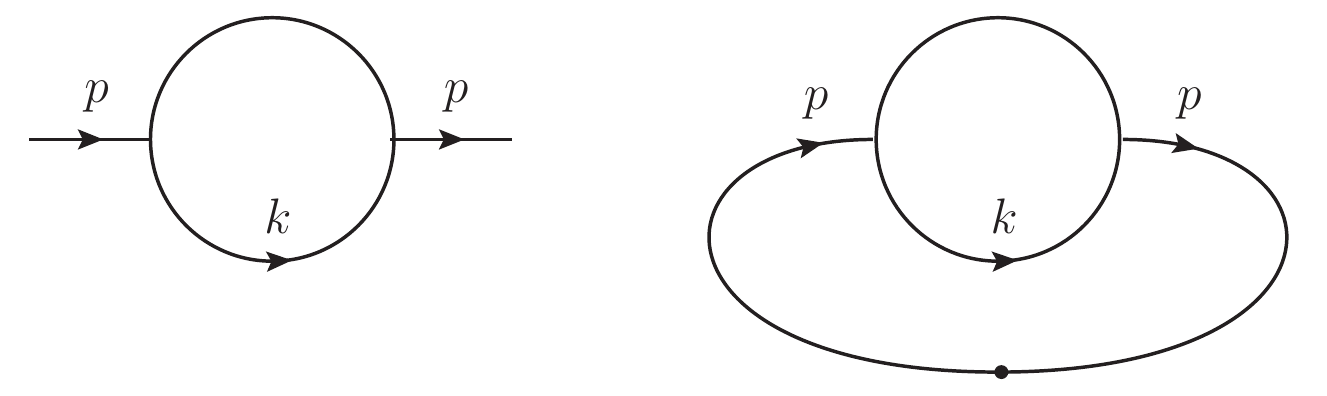}}
\caption{A Feynman diagram (left) and its adjoint (right)}
\label{adj}
\end{figure}

To say that a Feynman diagram $G$ is (non-)planar, one defines the \textit{adjoint} diagram $G^*$ (Fig.
\ref{adj}). It is constructed from $G$ by attaching all external lines to an auxiliary vertex
\cite{Nakanishi}.

We say that a Feynman diagram $G$ is planar iff $G^*$ is planar.
\section{Method I}
\label{sec:lapl}

Let $G$ be a connected $l-$loop Feynman diagram, $\{p_1,...,p_m\}$ be the set of external momenta, $\{e_1,...,e_n\}$ be the set of edges in $G$ (neglecting external lines), $\{k_1,...,k_n\}$ be the set of corresponding momentum flows in $G$ and $\{v_1,...,v_r\}$ be the set of vertices in $G$. Then there holds \cite{Weinzierl}
\begin{equation}\label{eqn:ver}
 r=n-l+1.
\end{equation}
Hence given the number of edges (flows) and loops, the number of vertices $r$ can be derived from (\ref{eqn:ver}). After introducing Feynman parameters $x_1,...,x_n$, one defines the Laplacian matrix $L$ of a graph $G$ as a $r\times r$ matrix with entries
\begin{align}\label{eqn:L1}
 L_{ij}=\left\{
\begin{array}{c l}     
    \sum\limits_{k=1}^{n} x_k & \mbox{if }i=j,\mbox{ }e_k\mbox{ is attached to }v_i,\mbox{ }e_k\mbox{ is not a self-loop,}\\
    -\sum\limits_{k=1}^n x_k & \mbox{if }i\neq j,\mbox{ }e_k\mbox{ connects }v_i,v_j.
\end{array}\right.
\end{align}
Elements of $L$ are calculated in a few steps. Diagonal elements $L_{ii}$ are obtained by deciding which
Feynman parameters $x_k$ are attached to $v_i$. Vertices are divided into external (attached to external
lines) and internal ones. Note that only triple and quartic (i.e. of degree 3 and 4) vertices are
allowed.\footnote{For more general applications like gravity, the algorithm should be improved. However, in
the case of $n-$ary vertices, the method II is a better approach.} Thus the conditions are of the form
\begin{equation*}
 \mbox{(for external vertices) }\pm k_a\pm k_b=\pm p_e\mbox{ or }\pm k_a\pm k_b\pm k_c =\pm p_e,
\end{equation*}
\begin{equation*}
 \mbox{(for internal vertices) }\pm k_a\pm k_b=\pm k_c\mbox{ or }\pm k_a\pm k_b\pm k_c=\pm k_d,
\end{equation*}
where $a,b,c,d\in\{1,...,n\}$, $e\in\{1,...,m\}$. The flows that fulfill the above relations contribute to
diagonal elements of $L$.
Furthermore, off-diagonal elements $L_{ij}$ are obtained by deciding Feynman parameters $x_k$ that connect
vertices $v_i$, $v_j$. Observe that such $x_k$'s have to be both in $L_{ii}$ and $L_{jj}$, thus the
intersection of elements in $L_{ii}$ and $L_{jj}$ is non-empty and gives exactly these $x_k$'s. In the case of
many edges connecting $v_i$, $v_j$, they shrink to one edge, thus giving exactly one $x_k$ (hence $\sum
x_k\rightarrow x_k$).\\
In order to obtain a more familiar form of $L$, understandable to \texttt{Mathematica} software \cite{math}, redefine $L$ by
\begin{align*}
 L_{ij}=\left\{
\begin{array}{c l}     
    \mathrm{deg}\left(v_i\right) & \mbox{if }i=j,\\
    -1 & \mbox{if }i\neq j\mbox{ and } v_i,v_j\mbox{ are adjacent},
\end{array}\right.
\end{align*}
where $\mathrm{deg}\left(v_i\right)$ is the degree of $v_i$. The above definition is derived from
(\ref{eqn:L1}) by substituting $x_k\rightarrow 1$. Then $L$ can be written as 
\begin{equation}\label{eqn:adj}
 L=D-A,
\end{equation}
 where $D=\mathrm{diag}\left(\mathrm{deg}\left(v_1\right),...,\mathrm{deg}\left(v_r\right)\right)$ is a degree
matrix and $A$ is the adjacency matrix given by
 \begin{align*}
 A_{ij}=\left\{
\begin{array}{c l}     
    1 & \mbox{if }i\neq j\mbox{ and } v_i,v_j\mbox{ are adjacent},\\
    0 & \mbox{otherwise}.    
\end{array}\right.
\end{align*}
The final part of the algorithm is to create the adjoint diagram $G^*$.\footnote{In the case of vacuum
diagrams, $G^*$ is the same as $G$ by definition.} The Laplacian matrix $L^*$ of $G^*$ is build upon $L$ by
extending it by one row and one column corresponding to $v_{r+1}$. Clearly
$\mathrm{deg}\left(v_{r+1}\right)=m$ and extra 1's appear in the $\left(r+1\right)^{th}$ column and
$\left(r+1\right)^{th}$ row at the elements corresponding to external vertices.  Thus, from (\ref{eqn:adj})
the adjacency matrix $A^*=D^*-L^*$ is obtained. Eventually, given $A^*$, the function \texttt{PlanarQ}
of the \texttt{Mathematica} package \texttt{Combinatorica} yields the answer for the question of
planarity of a Feynman diagram $G$. Additionally, in \cite{lee} there was made  a remark that it is
possible to draw a diagram upon the set of denominators. In fact, given the matrix $A^*$ it is possible
to draw a given diagram with \texttt{Mathematica} by using the function \texttt{AdjacencyGraph}.
Instructive examples for 
planarity recognition using the described algorithm are given in \cite{Katowice-CAS:2007}.

\section{Method II}
\label{sec:dual}
Let $G$ be a connected $l-$loop Feynman diagram, $\{p_1,...,p_m\}$ be the set of external momenta,
$\{k_1,...,k_n\}$ be the set of corresponding momentum flows in $G$. \textit{Dual variables}
$\{x_1,\ldots,x_m\}$ are defined by (see e.g. \cite{Arkani})
\begin{equation}
 p_1=x_1-x_m,\mbox{ } p_2=x_2-x_1,\mbox{     }\ldots\mbox{     },\mbox{ } p_m=x_m-x_{m-1}.
 \label{eqn:dual}
\end{equation}
Then by introducing rules of the form $k_i\rightarrow x_j\pm x_l$ all momenta are substituted by dual
variables and one obtains the following criterion: A Feynman diagram is planar iff it is possible to write all
propagators (including external momenta) in the form $x_i\pm x_l$. Let us present two examples.\footnote{The
following examples are massless, but the method is general and applicable also in massive cases, since masses
do not contribute to the momentum flow in a diagram.}

\begin{enumerate}
 \item Let $G$ be a simple box diagram with four external lines $\{p_1,\ldots,p_4\}$ and a loop momentum $k$. The amplitude is proportional to the integral
\begin{equation}
 \int\frac{\mathrm{d}^4 k}{k^2\left(k-p_1\right)^2\left(k-p_1-p_2\right)^2\left(k-p_1-p_2-p_3\right)^2}.
\end{equation}
Introduce dual variables according to (\ref{eqn:dual}).

\begin{figure}[t]
\centerline{%
\includegraphics[scale=0.65]{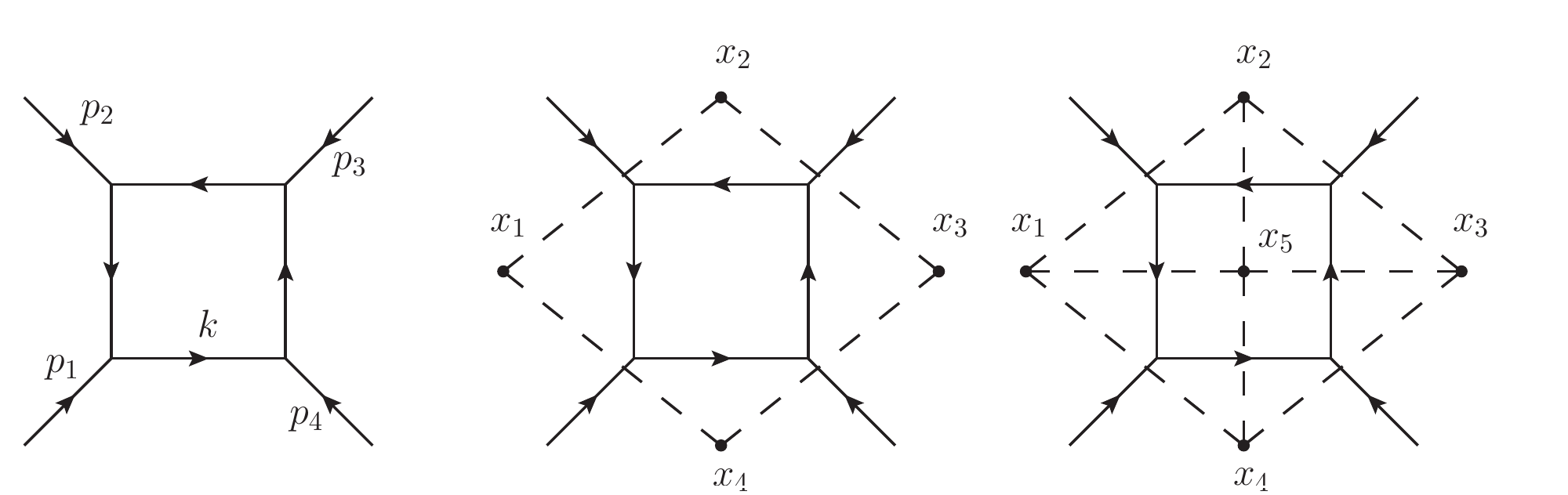}}
\caption{A box diagram, external dual variables and dual graph, respectively}
\label{adj1}
\end{figure}

Note that external momenta correspond to crossing of external lines and lines between corresponding dual variables (Fig. \ref{adj1}). The integral is now of the form
\begin{equation}
 \int\frac{\mathrm{d}^4 k}{k^2\left(k-x_1+x_4\right)^2\left(k-x_2+x_4\right)^2\left(k-x_3+x_4\right)^2}.
\end{equation}
Observe that substitution $k\rightarrow x_5-x_4$ gives a conformal invariant object
\begin{equation}
 \int\frac{\mathrm{d}^4 x_5}{\left(x_5-x_4\right)^2\left(x_5-x_1\right)^2\left(x_5-x_2\right)^2\left(x_5-x_3\right)^2}.
\end{equation}
\item Let $G$ be a non-planar double box with
\begin{equation}
 \int\frac{\mathrm{d}^4 k_1 \mathrm{d}^4
k_2}{
k_1^2\left(k_1-p_2\right)^2\left(k_1-p_1-p_2\right)^2k_2^2\left(k_2+p_3\right)^2\left(k_1-k_2\right)^2\left(k_
1-k_2+p_4\right)^2}
\end{equation}
and introduce dual variables again (Fig. \ref{bx}).

\begin{figure}[t]
\centerline{%
\includegraphics[scale=0.65]{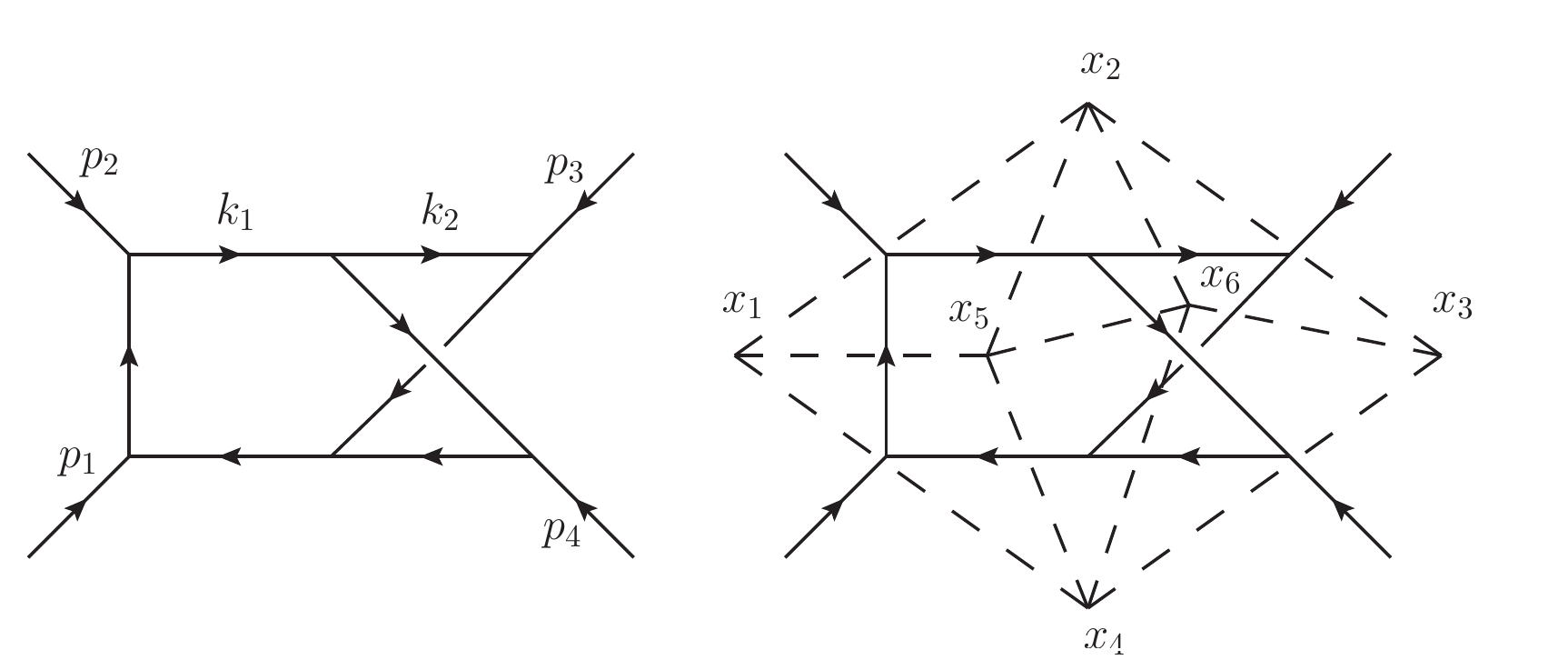}}
\caption{Non-planar double box (left) and its dual variables (right)}
\label{bx}
\end{figure}

Obviously, since the diagram is non-planar, it does not have a dual. Note that after transforming momenta to dual variables, one of the possible forms of the integral is

\begin{equation}
 \int\frac{\mathrm{d}^4 x_5 \mathrm{d}^4 x_6}{\left(x_5+x_2\right)^2\left(x_5+x_1\right)^2\cdot\ldots\cdot\left(x_6+x_5-x_3+x_4\right)^2}.
\end{equation}
\end{enumerate}

%\left(x_5+x_2\right)^2\left(x_5+x_1\right)^2\left(x_5+x_4\right)^2\left(x_6+x_3\right)^2\left(x_6+x_5\right)^2

Non-planarity is encoded in the element $x_6+x_5-x_3+x_4$, that breaks the conformal invariance. Thus there is a strict correspondence between dual diagrams and dual variables, hence another planarity criterion for Feynman diagrams is established. 
Instructive examples for planarity recognition using described algorithm are given in
\cite{Katowice-CAS:2007}. 

\section{Non-planar diagrams and dual variables}

\begin{figure}[htb]
\centerline{%
\includegraphics[scale=0.6]{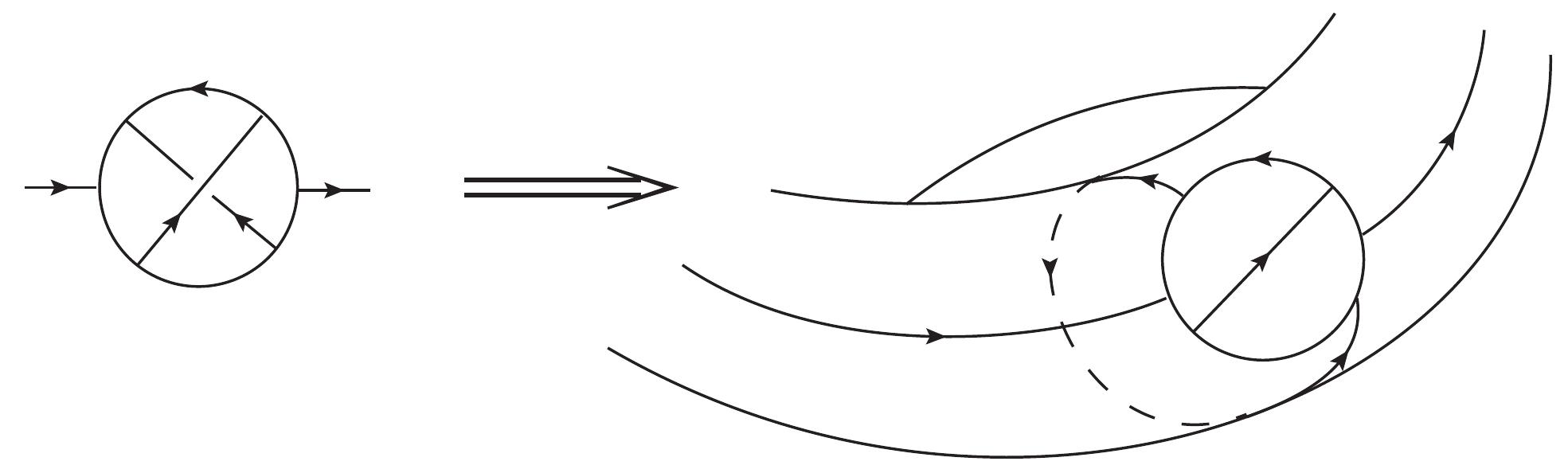}}
\caption{Three-loop non-planar self-energy (left) and its embedding on a torus (right)}
\label{tor}
\end{figure}

Observe that it is possible to planarize a non-planar diagram by embedding it on a surface with genus higher
than 0. Actually, the minimal genus of a surface where the given diagram $G$ is planar is called the
\textit{genus} of $G$. Let us give an example. Let a three-loop non-planar self-energy be embedded on a torus.
Then there is no crossing of diagram lines, hence the diagram is planar on the torus (Fig. \ref{tor}). It is
then possible to find its dual diagram and corresponding dual variables. Unfortunately, such an embedding sets
the number of faces too small to give a proper interpretation of momenta by dual variables. It can be easily
calculated by Euler's formula that
\begin{equation*}
  \chi=v-e+f=2-2g=2-2\cdot 1=0\longrightarrow f=0-6+9=3,
 \end{equation*}
 where $\chi$ --- Euler characteristics, $v$ --- number of vertices, $e$ --- number of edges, $f$ --- number of faces (dual variables), $g$ --- genus of a graph (surface). Thus there are only three dual variables $x_1$, $x_2$, $x_3$ available, in contrast to five momenta $p$, $-p$, $k_1$, $k_2$, $k_3$. Hence, although the non-planar diagram is planarized, it is not possible to define appropriate dual variables, since two momenta are absorbed by torus generators.

\label{sec:twist}

\section*{Acknowledgements}

Work supported by European Initial Training Network LHCPHENOnet PITN-GA-2010-264564. K. Bielas is supported by \'{S}wider PhD program, co-funded by the European
Social Fund.

\providecommand{\href}[2]{#2}

\begingroup\begin{thebibliography}{10}
\bibitem{ambre}
J.~Gluza, K.~Kajda, T.~Riemann, {\em Comput. Phys. Commun.} {\bf 177} (2007) 879; %\\, arXiv:0704.2423 [hep-ph];\\
  J.~Gluza, K.~Kajda, T.~Riemann and V.~Yundin,
  %``News on Ambre and CSectors,''
  Nucl.\ Phys.\ Proc.\ Suppl.\  {\bf 205} (2010) 147; ibid.
%  [arXiv:1006.4728 [hep-ph]];\\
 % J.~Gluza, K.~Kajda, T.~Riemann and V.~Yundin,
  %``Numerical Evaluation of Tensor Feynman Integrals in Euclidean Kinematics,''
  Eur.\ Phys.\ J.\ C {\bf 71} (2011) 1516.
%  [arXiv:1010.1667 [hep-ph]];

\bibitem{lee}
R. N.~Lee, {\em }{\bf }arXiv:1212.2685 [hep-ph].


\bibitem{Nakanishi}
N.~Nakanishi, {``Graph Theory and Feynman Integrals''} (Routledge 1971).
%%CITATION = HEP-PH 9905323;%%.

\bibitem{Arkani}
J.~Arkani-Hamed et al.,
%, J. L.~Bourjaily, F.~Cachazo, J.~Trnka, 
{\em JHEP} {\bf 1206} (2012) 125.%, arXiv:1012.6032 [hep-th].


\bibitem{Weinzierl}
C.~Bogner and S.~Weinzierl, {\em Int. J. Mod. Phys.} {\bf A25} (2010) 2585.%, arXiv:1002.3458v3 [hep-ph].

  
\bibitem{math} 
Wolfram Research, Inc., Mathematica, Champaign, IL (2010).



\bibitem{Katowice-CAS:2007}
Katowice, webpage http://prac.us.edu.pl/$\sim$gluza/ambre;\\
%\bibitem{Zeuthen-CAS:2004}
DESY, webpage http://www-zeuthen.desy.de/theory/research/CAS.html.



\end{thebibliography}\endgroup
\end{document}